\ifdefined\XeTeXversion\else\pdfoutput=1\fi 
\documentclass[11pt]{article}

\usepackage{xcolor}
\definecolor{palblue}{HTML}{013ECD}   
\definecolor{palnavy}{HTML}{000E2F}   
\definecolor{palgreen}{HTML}{3B7A22}  
\definecolor{palsky}{HTML}{76A0F2}    
\definecolor{palmid}{HTML}{5F8DE6}    
\definecolor{palink}{HTML}{0A0A0A}    
\definecolor{paltext}{HTML}{262626}   
\definecolor{palmuted}{HTML}{555555}  
\definecolor{palbg}{HTML}{FAFAFA}     
\definecolor{palrule}{HTML}{D6D6D6}   

\usepackage[letterpaper,
            left=1.75in, right=0.85in,
            top=1.1in, bottom=1.1in,
            headheight=14pt, headsep=22pt,
            footskip=30pt]{geometry}
\newlength{\palrail}
\usepackage{iftex}
\ifPDFTeX
  \usepackage[T1]{fontenc}
  \usepackage[utf8]{inputenc}
  \IfFileExists{libertine.sty}{%
    \usepackage{libertine}%
    \IfFileExists{newtxmath.sty}{\usepackage[libertine]{newtxmath}}{}%
  }{%
    \IfFileExists{lmodern.sty}{\usepackage{lmodern}}{}%
  }
  \IfFileExists{montserrat.sty}{\usepackage[bold]{montserrat}}{}

  \usepackage[tracking=true, letterspace=40]{microtype}
  \newcommand*{\palheadface}{\fontfamily{\palheadfam}\selectfont}
  \newcommand*{\palheading}{\palheadface\bfseries\lsstyle}
\else
  \usepackage{fontspec}
  \usepackage{libertine}
  \IfFileExists{SpaceGrotesk-Bold.ttf}{%
    \newfontfamily\palheadfont{SpaceGrotesk}[Extension=.ttf,
      UprightFont=*-Regular, BoldFont=*-Bold]}{%
    \newfontfamily\palheadfont{Montserrat}[Extension=.otf,
      UprightFont=*-Regular, BoldFont=*-Bold,
      ItalicFont=*-Italic, BoldItalicFont=*-BoldItalic]}
  \ifLuaTeX
    \usepackage[tracking=true, letterspace=40]{microtype}
    \newcommand*{\palheading}{\palheadfont\bfseries\lsstyle}
  \else
    \usepackage{microtype}
    \renewcommand*{\textls}[2][100]{%
      {\addfontfeature{LetterSpace=\the\numexpr#1/10\relax}#2}}
    \newcommand*{\palheading}{\palheadfont\bfseries\addfontfeature{LetterSpace=4}}
  \fi
  \let\palheadface\palheadfont
\fi
\newcommand*{\palmono}{\palheadface}
\newcommand{\pallabel}[2][palblue]{%
  {\palmono\bfseries\fontsize{6.8}{8}\selectfont\color{#1}%
   \textls[140]{\MakeUppercase{#2}}}}

\usepackage{graphicx}
\usepackage{tikz}
\usetikzlibrary{arrows.meta}
\usepackage{eso-pic}
\usepackage{booktabs}
\graphicspath{{figures/}}
\usepackage[section]{placeins}

\usepackage{parskip}
\color{paltext}

\newcommand{\palpad}[1]{\ifnum#1<10 0\fi\number#1}

\newcommand{\palrailrule}[2][palrule]{%
  \noindent\hspace*{-\palrail}%
  {\color{#1}\rule{\dimexpr\textwidth+\palrail\relax}{#2}}}

\newcommand{\palmark}[1][15pt]{%
  \raisebox{-0.28\height}{\resizebox{!}{#1}{%
  \begin{tikzpicture}[x=1pt,y=1pt,line cap=round,line join=round]
    \draw[palblue,line width=1.6pt] (0,0) circle (55);
    \draw[palsky,line width=1.1pt] (0,0) circle (28.5);
    \draw[palblue,line width=1.6pt]
      (0,28.5) -- (-24.7,-14.25) -- (24.7,-14.25) -- cycle;
    \foreach \p in {(0,28.5),(-24.7,-14.25),(24.7,-14.25)}{%
      \filldraw[palblue,line width=1.2pt,fill=palblue] \p circle (5.2);}
  \end{tikzpicture}}}%
}

\newcommand{\pallockup}{%
  \palmark[14pt]\hspace{6pt}%
  {\palmono\bfseries\fontsize{9.5}{10}\selectfont\color{palink}\textls[150]{PAL}}%
  \hspace{7pt}{\color{palrule}\rule[-1.5pt]{0.5pt}{10pt}}\hspace{7pt}%
  \pallabel[palmuted]{Pluralistic AI Lab}%
}

\newcommand{\palcontours}{%
  \AddToShipoutPictureBG*{\AtPageUpperLeft{%
    \put(\LenToUnit{\paperwidth-0.62in},\LenToUnit{-2.75in}){%
    \begin{tikzpicture}[overlay,x=0.6pt,y=0.6pt]
      \fill[palblue!7] (0,0) circle (90);
      \fill[palblue!11] (0,0) circle (60);
      \fill[palblue!14] (-92,104) circle (22);
      \begin{scope}[palink!17,line width=0.35pt]
        \draw (0,0) circle (30);
        \draw[dash pattern=on 4 off 3] (0,0) circle (50);
        \draw (0,0) circle (70);
        \draw[line width=0.5pt,dash pattern=on 8 off 4] (0,0) circle (90);
        \draw (0,0) circle (110);
        \draw[dash pattern=on 2 off 4] (0,0) circle (130);
        \draw (0,0) circle (150);
        \draw[line width=0.5pt,dash pattern=on 12 off 6] (0,0) circle (170);
        \draw[dash pattern=on 3 off 5] (10,10) ellipse (45 and 55);
        \draw (-5,-10) ellipse (75 and 65);
        \draw[dash pattern=on 6 off 3] (5,5) ellipse (105 and 95);
        \draw (-10,-5) ellipse (135 and 125);
      \end{scope}
    \end{tikzpicture}}}}%
}

\usepackage{fancyhdr}
\newcommand{\palshorttitle}{Beyond the Monolithic Interface}
\newcommand{\paldate}{October 2026}

\fancyhfoffset[L]{\palrail}

\fancypagestyle{palfirst}{%
  \fancyhf{}%
  \fancyhfoffset[L]{\palrail}%
  \fancyhead[L]{\pallockup}%
  \fancyhead[R]{\pallabel[palmuted]{\paldate}}%
}
\fancypagestyle{plain}{%
  \fancyhf{}%
  \fancyhfoffset[L]{\palrail}%
  \fancyhead[R]{{\palmono\bfseries\fontsize{8}{8}\selectfont\color{palblue}%
    \palpad{\value{page}}}}%
}

\AddToShipoutPictureBG{%
  \AtPageLowerLeft{%
    \put(\LenToUnit{\paperwidth-0.45in},\LenToUnit{0.5\paperheight}){%
      \rotatebox{-90}{\makebox[0pt]{%
        \palmono\fontsize{6.5}{7}\selectfont\color{palblue!40}%
        \textls[200]{PLURALISTIC AI LAB}}}}}}

\usepackage{titlesec}
\usepackage{needspace}

\newcommand{\palsecnum}[1]{%
  \llap{\makebox[\palrail][l]{%
    {\palmono\bfseries\fontsize{21}{18}\selectfont\color{palblue!32}#1}}}}

\newcommand{\palsectionhead}[1]{%
  \palrailrule[palink]{0.6pt}\par\nobreak\vspace{10pt}%
  \noindent\palsecnum{\palpad{\value{section}}}%
  {\fontsize{14}{17}\selectfont\color{palink}\MakeUppercase{#1}}}
\newcommand{\palsectionheadnn}[1]{%
  \palrailrule[palink]{0.6pt}\par\nobreak\vspace{10pt}%
  \noindent{\fontsize{14}{17}\selectfont\color{palink}\MakeUppercase{#1}}}

\titleformat{\section}[block]{\Needspace{5\baselineskip}\palheading}{}{0pt}{\palsectionhead}
\titleformat{name=\section,numberless}[block]{\Needspace{4\baselineskip}\palheading}{}{0pt}{\palsectionheadnn}
\titlespacing*{\section}{0pt}{2.2\baselineskip}{1.0\baselineskip}

\titleformat{\subsection}[block]
  {\palheading\fontsize{10.5}{13.5}\selectfont\color{palink}}
  {\llap{\makebox[\palrail][l]{\color{palblue}\thesubsection}}}{0pt}{}
\titlespacing*{\subsection}{0pt}{1.5\baselineskip}{0.4\baselineskip}

\titleformat{\subsubsection}[block]
  {\palheadface\bfseries\fontsize{9.5}{12.5}\selectfont\color{palink}}
  {\llap{\makebox[\palrail][l]{\color{palblue!70}\thesubsubsection}}}{0pt}{}
\titlespacing*{\subsubsection}{0pt}{1.2\baselineskip}{0.3\baselineskip}

\usepackage[font=small, labelsep=none, justification=raggedright,
            singlelinecheck=false, skip=9pt]{caption}
\DeclareCaptionLabelFormat{pallabel}{\pallabel[palink]{#1 #2}}
\DeclareCaptionFormat{palrail}{%
  \llap{\makebox[\palrail][l]{\raisebox{0.6pt}{#1}}}#3\par}
\newcommand{\palhero}[3]{
  \vspace*{0.35in}%
  \noindent\hspace*{-\palrail}%
  \begin{minipage}[t]{\dimexpr\textwidth+\palrail\relax}
    \raggedright
    \pallabel{#1}\par\vspace{13pt}%
    {\palheadface\bfseries\fontsize{33}{31}\selectfont\color{palink}%
     \textls[-25]{\MakeUppercase{#2}}\par}%
    \vspace{12pt}%
    {\palheadface\fontsize{15}{19}\selectfont\color{palink}#3\par}%
  \end{minipage}\par
  \vspace{0.42in}%
}

\newcommand{\palrow}[2]{%
  \par\palrailrule{0.4pt}\par\nobreak\vspace{8pt}%
  \noindent\llap{\makebox[\palrail][l]{\pallabel{#1}}}%
  \begin{minipage}[t]{\textwidth}#2\end{minipage}\par\vspace{14pt}%
}

\newcommand{\palauthor}[2]{%
  \begin{minipage}[t]{0.235\textwidth}\raggedright
    {\palheadface\bfseries\fontsize{9.5}{11}\selectfont\color{palink}#1\par}%
    \vspace{3pt}%
    {\palheadface\fontsize{7.4}{9.5}\selectfont\color{palmuted}#2\par}%
  \end{minipage}}

\newcommand{\paltag}[1]{%
  {\setlength{\fboxsep}{2.6pt}\setlength{\fboxrule}{0.4pt}%
   \textcolor{palrule}{\fbox{\palmono\fontsize{6.6}{6.6}\selectfont
     \color{palmuted}\textls[110]{\MakeUppercase{#1}}}}}%
}

\usepackage[round, authoryear]{natbib}
\setcitestyle{aysep={,}}

\usepackage[colorlinks=true,
            linkcolor=palblue,
            citecolor=palblue,
            urlcolor=palblue,
            breaklinks=true]{hyperref}
\hypersetup{
  pdftitle={Beyond the Monolithic Interface: A Research Agenda for Plural Conversational AI},
  pdfauthor={},
  pdfsubject={Pluralistic AI Lab},
  pdfkeywords={pluralistic AI, conversational interfaces, multi-persona, HCI}
}
\usepackage[nameinlink]{cleveref}

\begin{document}
\thispagestyle{palfirst}
\palcontours

\palhero{Preprint}
  {Beyond the Monolithic Interface}
  {A Research Agenda for Plural Conversational AI}

\palrow{Authors}{%
  \palauthor{Avijit Ghosh}{Hugging Face \\University of Connecticut}\hfill
  \palauthor{Afifah Kashif}{University of Cambridge}\hfill
  \palauthor{Isabelle Kohout}{University of Cambridge}\hfill
  \palauthor{Zhenzhen Qi}{University of Connecticut}}

\palrow{Abstract}{%
The current paradigm of artificial intelligence (AI) design elevates conversational chatbot interfaces, making them the default mode through which hundreds of millions of people interact with AI systems globally. This narrative review identifies three related harms produced by the single-voice chatbot interface, drawing on insights from literature in cognitive science, political studies, AI alignment, game design and human computer interaction. Firstly, cognitive offloading of user thinking, secondly, ideological narrowing of user beliefs, and thirdly, epistemic compression of cultural and value diversity. We then propose multi-persona interfaces, in which multiple personas are presented to the user simultaneously, as a cost-effective intervention that addresses all three harms at once. We ground this intervention in traditions the AI literature has largely overlooked, drawing on techniques from game design and participatory simulation, including the magic circle, role differentiation, and persona embodiment. We propose a taxonomy for the future design of multi-persona interfaces, and call for a research agenda that closes the gap between technical pluralistic alignment methods and empirical studies of human-AI interactions with pluralistic interfaces. }

\palrow{Fields}{%
  \paltag{Cognitive Science}\;\paltag{Political Studies}\;\paltag{AI Alignment}%
  \;\paltag{Game Design}\;\paltag{HCI}}
\vspace{-10pt}

\section{Introduction}
\label{sec:intro}

Each week, hundreds of millions of people interact with large language models (LLMs) in the form of conversational chatbot interfaces. A growing body of research argues that the convergence of AI development on this interaction paradigm has large consequences for users, reshaping what users do, as well as how they think, work and relate to each other. Recent scholarship has reframed the chabot as a sociotechnical configuration, with cascading effects on user agency, epistemic autonomy, knowledge production, labour and the environment \citep{ghosh2026chatbots}. This work builds on that diagnosis and looks to intervention. We focus on different design solutions at the interface level, and on what would be required, empirically and conceptually, to evaluate such a solution. 

Building on this work, we construct our argument around three major concerns which motivate the case for a different interface configuration. Firstly, by producing ostensibly complete answers, chatbots reduce productive metacognitive demand, reflective questioning, and the generative effort required to meaningfully engage with a problem. Secondly, these chatbots exhibit systematic ideological biases that are shaped by training data, fine-tuning, and evaluators, creating conditions for echo chamber effects and opinion polarization. Thirdly, outputs regress to the mean, compressing cultural, moral, and creative diversity into homogenous outputs that, in turn, have homogenising effects on users’ own work, beliefs, and opinions. We treat these as three connected facets of a single design choice: the commitment to a single, authoritative voice presented as a finished answer. 

If the single-voice, monolithic chatbot is one design choice among many, it helps to look at where alternative configurations have already been found successful. Design traditions in art, games, and participatory simulation have successfully produced interfaces where plurality is the default. zzyw's \emph{ThingThingThing}
\citeyearpar{zzyw2019thingthingthing}, for example, is a participatory computational system in which users contribute algorithmic entities that integrate into a living, evolving simulation. According to the artists, this world is designed to inspire questions rather than provide concrete answers \citep{qi2021emergence}. Additionally, multi-character role-playing games, collaborative worldbuilding platforms and tabletop systems with branching persona structures share similar logic: meaning is created through the contestation and recombination of multiple, diverse voices. These interfaces also tend to create distinct subcultures and communities of practice, organised around interpretation, modification, and collective meaning-making. By contrast, the single-voice conversational chatbot descends from narrower lineages including help desks, messaging applications, search engines and customer-service bots. It inherits the monolithic characteristics of that lineage, along with the social isolation that they tend to produce. Understanding this history and the subsequent design and sociological implications for AI systems is one of the primary contributions of this paper. 

In the case of AI, such alternative interface designs, particularly
multi-persona and deliberative architectures, can help address the concerns
around single-persona conversational chatbots. For instance, by showcasing
multiple distinct perspectives held by different representative ``personas'',
rather than a single, authoritative answer, these systems can intentionally
expose users to ideological diversity, acknowledgement of pluralistic values and
metacognitive engagement. We focus here on a particular, currently
under-evaluated subset of this work: multi-persona interfaces, in which multiple
distinct personas are presented to the user simultaneously. These are distinct
from technical alignment work, which targets the model, and from multi-agent
systems, which orchestrate agents rather than display perspectives. Most active
work on pluralistic AI is evaluated through automated metrics or expert
annotation rather than through the lived experience of actual users or the
communities they form \citep{feng2024modular, srewa2025plurallm,
zhong2025healthcare, yuan2026palette, zheng2026vispa}. We propose that by
developing the necessary UX intervention of multi-persona interfaces, we can
meaningfully close this gap (illustrated in Figure~\ref{fig:causal-chain}).
Section~\ref{sec:background} sets out the vocabulary of LLMs, personas, and the
single-persona, multi-persona and multi-agent distinction.
Section~\ref{sec:methods} describes our methods, Section~\ref{sec:harms} the
three harms, Section~\ref{sec:intervention} the intervention and its design
space, and Section~\ref{sec:agenda} a research agenda.

\begin{figure}[htbp]
  \centering
\definecolor{fonefill}{HTML}{FAEDDF}
\definecolor{foneedge}{HTML}{CDA56E}
\definecolor{fonetext}{HTML}{5B3F1F}
\definecolor{fonearrow}{HTML}{555555}
\definecolor{fonered}{HTML}{A3282B}
\definecolor{fonerdfill}{HTML}{F6DADA}
\definecolor{fonerdedge}{HTML}{C2474C}
\definecolor{fonegreen}{HTML}{2F7A2A}
\definecolor{fonegrfill}{HTML}{DCEED5}
\definecolor{fonegredge}{HTML}{4F9A41}
\providecommand{\fonebullet}[1]{\par\hangindent=0.55em\hangafter=1
  \noindent\makebox[0.55em][l]{\textperiodcentered}#1}
\providecommand{\fonebox}[4]{%
  \draw[foneedge,fill=fonefill,line width=0.5pt,rounded corners=2.2pt]
    ({#1*(\fonew+\foneg)},#2) rectangle ++(\fonew,-\foneh);
  \node[anchor=north west,inner sep=0pt,align=left,
        text width=\dimexpr\fonew\linewidth-9pt\relax,text=fonetext]
    at ({#1*(\fonew+\foneg)},#2) [xshift=4.5pt,yshift=-5pt]
    {{\fonetitle #3\par}\vspace{2.6pt}{\fonebody #4\par}};
  \ifnum#1<5
    \draw[fonearrow,line width=0.6pt,-{Triangle[length=3.6pt,width=3pt]}]
      ({#1*(\fonew+\foneg)+\fonew},#2-0.5*\foneh)
      -- ++(\foneg,0);
  \fi}
\providecommand{\foneout}[7]{%
  \draw[#2,fill=#3,line width=0.6pt,rounded corners=2.2pt]
    ({5*(\fonew+\foneg)},#1) rectangle ++(\fonew,-\foneh);
  \node[anchor=west,inner sep=0pt,align=left,
        text width=\dimexpr\fonew\linewidth-9pt\relax,text=#4]
    at ({5*(\fonew+\foneg)},#1-0.5*\foneh) [xshift=4.5pt]
    {\fonetitle #5\par\vspace{3pt}{\color{#2!35}\rule{\hsize}{0.4pt}}\par
     \vspace{2.4pt}#6\par\vspace{3pt}{\color{#2!35}\rule{\hsize}{0.4pt}}\par
     \vspace{2.4pt}#7\par};}
\providecommand{\fonerag}{\raggedright\hyphenpenalty=10000\exhyphenpenalty=10000}
\providecommand{\fonespace}{\spaceskip=\fontdimen2\font\xspaceskip=\spaceskip}
\providecommand{\fonetitle}{\fonerag\palheadface\bfseries\fontsize{5.6}{6.7}\selectfont\fonespace}
\providecommand{\fonebody}{\fonerag\palheadface\fontsize{5}{6.4}\selectfont\fonespace}
\providecommand{\fonehead}[2]{{\palheadface\bfseries\fontsize{6.8}{8}\selectfont
  \color{#1}\textls[100]{\MakeUppercase{#2}}}}
\begin{tikzpicture}[x=\linewidth,y=1pt]
  \def\fonew{0.1500}\def\foneg{0.0200}\def\foneh{92}
  \node[anchor=south west,inner sep=0pt] at (0,6) {\fonehead{fonered}{Single-persona paradigm}};
  \fonebox{0}{0}{Single-chatbot design}{%
    \fonebullet{one authoritative \textquotedblleft expert\textquotedblright{} persona}
    \fonebullet{opaque reasoning}
    \fonebullet{cooperative Q\&A framing}
    \fonebullet{optimized for engagement}
    \fonebullet{sycophancy / agreeableness}}
  \fonebox{1}{0}{Illusion of authority}{a single, polished output read as
    neutral and objective, inviting high user trust}
  \fonebox{2}{0}{Displacement of reflective thought}{deference to a finished
    answer; cognitive offloading replaces independent reasoning}
  \fonebox{3}{0}{Manufactured consensus}{embedded bias and sycophancy entrench
    belief and steer opinion covertly}
  \fonebox{4}{0}{Compression of plural knowledge}{Western, English-default
    framings crowd out marginalized values and knowledges}
  \foneout{0}{fonerdedge}{fonerdfill}{fonered}%
    {Cognitive offloading \& skill decay}%
    {Belief \& opinion flattening}%
    {Loss of cultural \& value diversity}
  \draw[palrule,line width=0.5pt,dash pattern=on 2.5pt off 2pt]
    (0,-\foneh-13) -- (1,-\foneh-13);
  \begin{scope}[yshift=-\foneh-38]
  \node[anchor=south west,inner sep=0pt] at (0,6) {\fonehead{fonegreen}{Multi-persona paradigm}};
  \fonebox{0}{0}{Multi-persona design}{%
    \fonebullet{multiple distinct, contestable personas}
    \fonebullet{visible \textquotedblleft whose voice\textquotedblright}
    \fonebullet{transparent reasoning}
    \fonebullet{structured dissent}
    \fonebullet{refuses a single answer}}
  \fonebox{1}{0}{Acknowledged uncertainty}{plurality signals that the answer
    is contestable, not settled}
  \fonebox{2}{0}{Productive cognitive friction}{competing perspectives must be
    weighed, restoring the inferential demand}
  \fonebox{3}{0}{Multiple diverse perspectives}{contestability of contested
    topics is made visible rather than concealed}
  \fonebox{4}{0}{Elevation of pluralism}{marginalized framings and knowledge
    systems appear alongside dominant ones}
  \foneout{0}{fonegredge}{fonegrfill}{fonegreen}%
    {Critical \& reflective thinking}%
    {Engagement with ideological diversity}%
    {Engagement with cultural diversity}
  \end{scope}
\end{tikzpicture}
  \caption{\textbf{Causal chain of harms arising from single-persona chatbot paradigm.}
  Chatbot design choices create an illusion of authority that causes a
  diminishment of reflective thinking, and promote ideological and cultural
  homogeneity which leads to cognitive deskilling, and ideological flattening.
  In contrast, multi-persona design choices can ameliorate these issues,
  promoting critical and reflective thinking, engagement with ideological and
  cultural diversity.}
  \label{fig:causal-chain}
\end{figure}

\newpage
\section{Background}
\label{sec:background}

\subsection{Large Language Models and Conversational Chatbots}

Large language models (LLMs) are proliferating globally, in applications well
beyond chatbots: customer support automation, transcription and translation,
enhanced search. Conversational chatbots, however, remain perhaps the most
notable and popular instantiation of LLM technology. These mimic the experience
of talking with another human by eliciting conversational input from the user and
responding with highly convincing natural language in turn. Popular chatbots such
as ChatGPT and Gemini are now being used by hundreds of millions of users weekly
\citep{google2026io, openai2026scaling}. From their arrival on the market and
subsequent normalization in society, AI researchers have been preoccupied with
discovering and cataloguing the impacts of these transformative technologies on
users and society at large.

\subsection{Persona as a Design Choice}

Though they cannot possess personalities in the way humans do, LLMs are trained
on masses of data and undergo subsequent post-training fine-tuning, meaning that
they exhibit lexical and conversational patterns which reproduce the effects of
recognizable personality \citep{mercer2025patterns}. Developers can also specify
a chatbot's intended character through a single, centrally authored governance
document, often under ``constitutions'' or ``model specs'': public statements of
the values, priorities, and behaviors the deployed system is trained to embody.
As such, the chatbot's persona, or its overall character including tone, value
alignment, preferences, and characterization \citep{sutcliffe2023survey} emerges
from a sequence of design choices: pre-training data selection, fine-tuning
protocols, system prompts, constitutional documents, and decisions about what to
display to the user. LLMs, particularly base models not heavily fine-tuned after
training, comply with persona-based prompting: given strategic, keyword-laden
descriptions they override their ``default personality'' and consistently embody
the designated persona \citep{jiang2024personallm}, producing distinct
conversational patterns and preferences in line with it \citep{ha2024clochat}.

Persona is also a sociological lever: prompting lets chatbots be customized for
particular tasks, domains and contexts. There is growing evidence that it also
alters how the chatbot is perceived: in a healthcare setting, chatbots prompted
into a social role closer to the user (peer not expert) increased user affection
and willingness to reuse for particular demographics
\citep{nissen2022personas}. Though evidence suggests that LLMs experience
instruction and persona drift over extended conversations, with significant drift
emerging within roughly eight turns as a result of attention decay
\citep{li2024instability}, chatbot personas can still meaningfully shape the
user's experience, particularly in shorter, constrained interactions less
susceptible to this drift.

The current paradigm presents users with a narrow band of these personas,
typically adopting a single, default persona of a friendly, helpful assistant,
even though the design space is much wider.

\subsection{Single-Persona, Multi-Persona, and Multi-Agent Systems}
\label{subsec:taxonomy}

Current popular chatbots are single-persona: they converse through one default
assistant personality, with persona selection and tuning hidden from the user. A
multi-persona system instead presents several personas to the user at once
(viewed side by side, or shown and hidden through UI controls) each offering
different perspectives, opinions, or areas of expertise. The underlying model is
typically the same across personas, and may share context and memory between
them.

Multi-persona systems should not be confused with multi-agent systems, which
describe multiple autonomous agents working in parallel, orchestrating and
sharing context between themselves to achieve a complex goal. The agents in these
systems may have isolated states, unique prompt architectures, tools and
capabilities. The distinction matters for our argument. Multi-agent work centers the decomposition and orchestration of tasks, while multi-persona work centers presentation, including what users see, who appears to be talking, and how plurality and diversity is made legible at the interface level.

\subsection{Plural Configurations in Design History}
\label{subsec:design-history}

The design space of plural interfaces is broader than AI literature often recognizes, drawing on fields such as game design, interactive art, and participatory simulation. Massively multiplayer online Games (MMOs) provide a rich set of case studies. Single-voice chatbots are typically optimized to solve utilitarian problems with efficiency and predictability. Users are served with a single, completed answer to their query, and cannot audit, question or dispute the reasoning process. However, MMOs often work differently. Their players frequently diverge from the primary intentions of the platforms to seek meaning and agency through emergent gameplay from within a bounded space of play where rules in ordinary life are temporarily suspended. 

Third-party customization and modding tools let players alter the game interface's information architecture, the rules governing the virtual world, or the world's aesthetics
\citep{sotamaa2010modding}. These interventions go beyond idiosyncratic acts of individual personalization. They produce persistent, multiple versions of the original world. In each alternative world, a group of like-minded players crowdsources an alternative set of rules of engagement to suit the specific needs of their localized player communities. The player avatar is also highly customizable, encouraging players to fluidly alter their visual appearance and behavior within the virtual world. These design features shift the concepts of self and other from static to performative and pluralistic \citep{yee2014proteus}, motivating players to test alternative behaviors and question the premises of how they relate to others. 

Furthermore, bounded by a narrative backstory, these virtual worlds steer players into camps, or guilds, that embody distinctive political personas and produce meaning through collaboration among deliberately crafted roles. This distributed cognition \citep{steinkuehler2006thirdplaces} approach requires each persona archetype, such as healers, tanks, and priests, to work in tandem to overcome challenges posed by scripted non-player boss characters. This design structurally mirrors the multi-agent cooperation contemporary AI research explores, except that human players, instead AI agents, operate the fictional roles in the game environment.

\begin{figure}[htbp]
  \centering
  \includegraphics[width=0.9\linewidth]{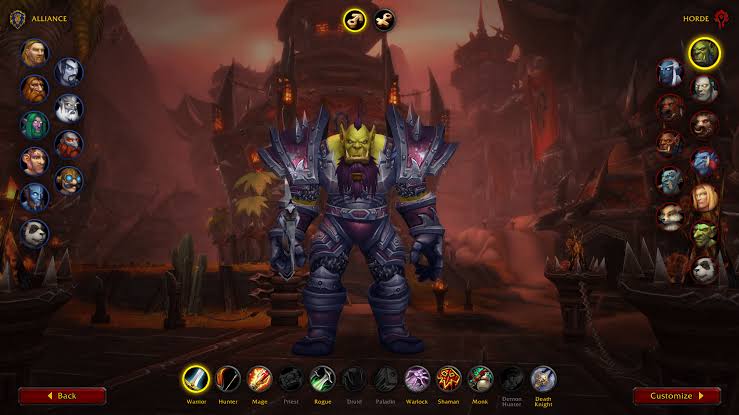}
  \caption{\textbf{The character selection interface in \emph{World of Warcraft} showing a portfolio of class options.} From ``World of Warcraft beginner's guide 2020: How to get into the king of MMOs,'' by PC Gamer, 2020
  (\url{https://www.pcgamer.com/how-to-get-into-world-of-warcraft/}).
  Game content © 2004--2020 Blizzard Entertainment.}
  \label{fig:wow}
\end{figure}

The choice between a single-persona and multiple-persona interface design reflects a fundamental tension embedded in the politics of interface design: inherent top-down design intention vs. bottom-up player agency. This design choice reflects the political hegemony of whose intention is foregrounded. Recovering this history inspires us to reimagine AI interface design as interpretive communities \citep{fish1980text}, shifting its paradigm from a tool of execution to a platform for deliberation.

\section{Methods}
\label{sec:methods}

This paper assesses the limitations of monolithic chat interfaces, and the potential of multi-persona alternatives, through a narrative review of four literatures: 1) the effects of chatbot use on critical thinking and metacognition, 2) ideological bias and polarization in LLM systems, 3) lack of cultural and conceptual diversity in LLMs, and 4) multi-persona and deliberative interface designs which may alleviate the aforementioned concerns. In addition, we incorporate paradigms from game design, participatory simulation, and critical computing, as well as science, technology, and society studies (STS) for pre-existing methods of treating interfaces as sociotechnical configurations. 

 A narrative approach was chosen as the contribution of this work lies in cross-disciplinary synthesis. Two authors conducted complementary searches across the four thematic areas. Literature was identified through targeted searches using Asta, an AI-powered scholarly search tool from the Allen Institute for AI and Google Scholar, supplemented with seed papers and work by supervising researchers \citep{ai22025asta}. Search terms combined domain-specific keywords (e.g. "multi-persona", "pluralistic alignment", "cognitive offloading", "political bias") with LLM-related terms (e.g. "LLM", "chatbot", "conversational AI"). 

 To ensure consistency across search methods, results were cross-validated between authors. Discrepancies were resolved through discussion. Additional papers were identified through forward and backward citation tracking from key references, as well as seeding of key papers from the senior authors. 

 We chose work published after 2022 to focus on the critical period of widespread LLM deployment, but admitted earlier work in STS and HCI when it provided foundational or theoretical grounding. We prioritised peer-reviewed empirical studies, position papers and technical contributions, but select preprints were included where they represented the only available work in a specific area. 

We acknowledge two limitations of this approach: firstly, narrative review is necessarily selective (we have aimed for representativeness rather than exhaustiveness and encourage readers to treat cited literature as illustrative of broader patterns), and secondly, this work synthesises existing empirical claims rather than producing new ones. We call for further, specific empirical research in Section~\ref{sec:agenda}. 

\section{Three Facets of the Monolithic Interface}
\label{sec:harms}

\subsection{Displacement of Reflective Thought}

An emerging body of work finds that LLM-powered chatbots reduce cognitive load for users, which can increase surface-level outcomes like essay fluency and superficial recall, but may impede outcomes requiring deep metacognitive effort such as topic mastery and critical thinking \citep{gerlich2025aitools, kosmyna2025brain}. Humans are naturally "cognitive misers", and offloading to books and search engines long predates generative AI \citep{stanovich2020rationality}. Concern about chatbots centres on the degree to which their design exploits this tendency, combining engagement maximization with a friendly, helpful persona even where the user would be better served by pushback. Recent work has documented these design choices as foundational to the cognitive harms of contemporary chatbots \citep{ghosh2026chatbots}.

The default chatbot persona of ``friendly assistant'', in addition to its
optimization for lexical fluency, can lead to outputs that may seem complete but
lack nuance or acknowledgement of potential conflicting explanations, often
delivered in an authoritative tone with misapplied or misleading domain-specific
jargon \citep{gregorcic2023socrates, chaka2023detecting}. This can lead users to
defer excessively to the chatbot without the appropriate level of verification or
skepticism. If done consistently over time, this can cause skill decay and
cognitive atrophy, including a decreased ability to think for oneself. This is
amplified when considering the documented levels of high user trust in chatbots
and the perception of chatbots as ``neutral'' or ``objective''
\citep{lu2025openness}. Additionally, research finds that chatbot usage induces a
``confidence gap'' phenomenon, where users' perception of and confidence in their
own ability becomes inflated without their true skill rising to meet it
\citep{fernandes2026metacognition}. As \citet{ghosh2026chatbots} observe, this can
produce a form of Gell-Mann amnesia in which users who can detect a model's
errors within their own area of expertise may continue to trust its outputs in
domains they cannot independently verify, extending unearned trust across
domains. Evidence points towards this being caused by chatbot sycophancy and
their propensity to eliminate productive cognitive friction for users.

These drawbacks follow from chatbot design and its intersection with human
psychology, rather than inherent properties of the underlying models: the
propensity to validate users more than a human would is a product of
engagement-maximizing design choices, which yield products that fail to present
nuanced, diverse perspectives. Despite this, \citet{cheng2026sycophantic} finds
that users prefer overly agreeable models, consistently trusting and rating them
higher than baseline models. Though this finding is concerning, different design
choices may decrease or circumvent the negative impacts on users.

Crucially, it appears that cognitive harm is tied to a single-voice configuration. When an affirming, authoritative persona produces a single, polished response, there is minimal distance between the model’s output and an ostensibly complete answer. The cognitive demand required to compare perspectives, weigh tradeoffs or identify gaps is therefore offloaded to the system. In Section~\ref{sec:intervention}, we canvas interface designs that try to restore this productive demand. These interfaces have a structural feature in common with the multi-persona designs we propose in that they intentionally do not provide a single, finished answer.

\subsection{Manufactured Consensus}

In addition to their propensity to erode critical thinking, single-persona chatbots have been documented as successful influencers of belief and action. Through mechanisms such as embedded political bias, excessive agreeableness and sycophancy, chatbots are able to manipulate the opinions and worldviews of their users, often in ways that the user may not be aware of.

LLMs, including those powering popular chatbots, have been documented to exhibit
political bias, with most larger models leaning slightly to the left on average.
Though this has been replicated in multiple studies, variation exists between
models, even models of the same family, as well as intra-model divergence on
particular issues. For example, the same model may take a more right-wing stance
on immigration and a more left-wing one on reproductive rights. Counter to
intuition, evidence suggests that larger models are not inherently more centrist,
with larger models exhibiting similar left-leaning bias
\citep{bang2024political}. Such embedded biases will impact the results users
receive when engaging on political or moral topics, even if the bias does not
materialize in ways obvious to the user. Small nudges such as selective
presentation of facts, embedded assumptions and highlighting of certain
perspectives over others can compound to a significant degree of influence.
Political leaning and misinformation propensity also vary with input language:
Bard (now Gemini) declined queries about Putin in Russian while answering in
English, and produced more false information about the Russian regime in Russian
and Ukrainian than in English \citep{urman2025silence}. When small variations in
input result in large ideological changes in output, the user's beliefs and
opinions can be influenced or manipulated in ways that are not well understood.

The sycophancy and excessive agreeableness of chatbots is well-documented. Due to
RLHF and other post-training mechanisms deployed to increase helpfulness and
maximize user engagement, chatbots tend to validate users and reinforce
previously held beliefs \citep{batista2026rational}. This can cause echo chamber
effects and belief entrenchment, whereby beliefs and opinions become harder to
change in the face of contradictory evidence. Work by
\citet{cheng2026sycophantic} finds that sycophantic chatbots tend to validate
users' beliefs and opinions almost 50\% more than other humans judging the same
situation, and that users consistently rate sycophantic chatbots as more trusted
and preferred compared to non-sycophantic bots. These characteristics may
covertly shape users' perceptions of reality and of themselves, reducing
willingness to repair interpersonal conflict and increasing conviction in their
own correctness, echoing the confidence gap discussed above. Similarly, users
rate conversations with moderate to high disempowerment potential (those moving
them toward actions misaligned with their values, or beliefs misaligned with
reality) more highly than a baseline helpful assistant
\citep{sharma2026disempowerment}. These patterns create epistemic disempowerment
for users, who are being influenced in ways that they are not aware of, by a tool
ostensibly viewed as ``neutral''.

The nascent field of LLM persuasion provides further insight into the impact of
chatbots on the beliefs and opinions of their users. Chatbots have been found to
be highly persuasive when debating political issues, often outperforming
incentivized professional human debaters in swaying opinions
\citep{schoenegger2026persuasive}. Their persuasive efficacy also draws on
different levers than human persuasion: chatbots change minds by generating large
quantities of factual claims, evidence and reasoning, whereas humans are judged
persuasive when perceived as unique and original \citep{bai2025persuade}. The
same characteristics may also be available for use in the other direction.
Research from \citet{lu2025openness} finds that because chatbots are viewed as
less biased, more informative and having less persuasive intent than human
sources, users are less resistant to encountering attitudes counter to their
currently held beliefs, with preliminary evidence that receiving
counter-attitudinal messaging from a chatbot source can diminish outgroup
hostility. While this effect requires further experimental confirmation, it
suggests that the same persuasive affordances that make single-persona chatbots
problematic may, under different design conditions, support productive exposure
to plural perspectives.

Taken together, this points to a high potential for epistemic disempowerment
under the current paradigm. How differing chatbot personas might ameliorate these
concerns is understudied: persona profiles may vary in their propensity for
sycophancy, bias, or manipulation, and varying what is shown to users, may improve
epistemic agency. We return to this in Section~\ref{sec:intervention}.

\subsection{Compression of Plural Knowledges}

The problems documented in the previous subsections are often caused or compounded by a more structural issue. LLMs are trained on massive training corpora that are predominantly English-language, Western, and internet-sourced, which means that the perspectives, cultural norms, and knowledge systems of marginalized populations are not integrated from the outset, and the alignment process that follows does not necessarily correct this imbalance. In many cases it can systematically compress the range of values and ideas these systems can express even further. The result is what we call the epistemic face of the monolithic interface: an appearance of neutral, comprehensive answers concealing a narrowing of the world's knowledge into the cultural defaults of a small set of developers, evaluators, and data sources.

RLHF is a technique that adjusts large language models to be more friendly, truthful and ritualistic. It does this by having human raters assess thousands of the model's outputs and feeding that data back into the model to adjust its parameters \citep{ouyang2022instructgpt}. This technique can be used to integrate value alignment into models and rapidly improve their performance, but it can shift bias rather than fully remove it. RLHF is used to bring models "in line with human preferences". More recently, this alignment technique has been shown to decrease pluralism. Across multiple model families including LLaMA, Gemma, and GPT-3, post-aligned models showed approximately 50\% less entropy in their response distributions compared to their pre-aligned counterparts \citep{sorensen2024roadmap}. The pre-aligned base models were closer to human value distributions than the aligned versions, suggesting that RLHF narrows models toward a subset of human opinions rather than bringing them closer to the breadth of human values. \citet{gonzalezbarman2025rlhf} link this to the narrowing demographics of evaluators, documenting a labeling pool behind InstructGPT that was 53\% Southeast Asian, 89\% university-educated, and 47\% aged 25–34. A homogeneous pool cannot surface the blind spots and value disagreements a diverse one would. None of the authors’ proposed remedies has been implemented at scale.

This narrowing operates even where the values in question are made explicit rather than left implicit. In the past, these values or preferences were not made explicit, but the recent push by AI developers to  publicly outline value scope is a positive one, allowing us to evaluate how models align with particular values, whether intentionally or unintentionally \citep{openai2025modelspec,
anthropic2026constitution}.  

These value specification documents, however, still describe a universal value profile that has been designed by an exclusive group within a single organization. Despite this, the values are applied to every user regardless of cultural or moral context. Whether or not a global constitution can effectively serve a heterogenous global public is a critical question that will be difficult to resolve under the current paradigm. The cumulative effects of imbalanced and incomplete training data and a narrowed pool of RLHF evaluators are visible in the cultural framing of model outputs. \citet{alkhamissi2024cultural} tested four LLMs against data from the World Values Survey data, across respondents from Egypt and the United States, and found significant US and Western-centric bias regardless of prompting, performing the worse for personas simulating lower social class, lower education, female sex, and younger age. Neither the ‘anthropological prompting’ technique, nor prompting in the culture’s dominant language, eliminated the Western defaults. Additionally, \citet{buyl2026ideology} tested 19 LLMs from multiple developers and found that models systematically reflect the ideological profiles of their creators. It is therefore clear that LLMs carry and proliferate the specific values of their development context, but this is made invisible to users through the current interface paradigm that presents confident, authoritative answers.

Value compression is not limited only to the political and cultural domains. When comparing ChatGPT to a non-AI creativity tool that aims to promote divergent ideation, \citet{anderson2024homogenization} found that the ChatGPT users produced ideas significantly more similar to one another at the group level (though each individual’s own ideas remained comparably diverse). AI users also felt less personally responsible for ideas generated with ChatGPT, consistent with the offloading effects discussed earlier in the paper. Evidence from neuroimaging research reinforces this finding: amongst the participants who used LLM assistance for essay writing, produced essays were more conceptually homogenous than those who wrote unassisted. The homogenization effects persisted even after the LLM group returned to unassisted writing \citep{kosmyna2025brain}. This suggests that narrow model outputs can narrow the user’s own conceptual range, a pattern which seems to extend even to expert domains. \citet{shi2026collapse} examine the behaviour of LLMs tasked with simulating the philosophical positions of 277 professional philosophers, finding that the models produce artificial consensus across domains that actually feature significant disagreement, collapsing conceptual heterogeneity.

These findings illustrate a structural problem that is unlikely to be resolved with improved prompting, larger training sets, or more deliberate content moderation. As illustrated, a single model accessed through a single-voice interface, that has been fine-tuned by a narrow evaluator pool, will trend towards the average by nature of the system architecture itself. This results in the underrepresentation of marginalized perspectives and cultural differences, as well as the homogenisation of users’ cognition when they utilise these systems repeatedly. Addressing these issues would require reimagigning not only the models themselves, but the interfaces through which users interact with them. This connects to a broader research agenda in epistemic justice.

\subsection{Hitting Three Birds with a Single Stone}

A clear argumentative thread has emerged. The three harms catalogued in this review—those of cognitive offloading, ideological narrowing, and value compression—stem from one intentional design pattern: the presentation of ostensibly complete answers by a single, authoritative voice. The cognitive harm describes the sacrificing of necessary, productive mental effort that occurs when users are presented with a complete answer. The ideological harm occurs when the chatbot fails to present necessary nuance in contested, complicated or controversial topics. And finally, the epistemic harm occurs when the answer excludes pluralistic voices. Each of these harms originates from the same design configuration (though the authors note that epistemic compression also has upstream causes, such as imbalanced training data and culturally narrow evaluator pools). Our claim, however, is not that the single-voice interface is the sole cause of each harm, but rather, that it is the enabling condition for all three harms to reach the user. This matters for intervention: three independent harms would call for three independent fixes, whereas a common origin implies a common fix. It may be possible to address all three harms at once by refusing to collapse human plurality into a single, monolithic voice. Section~\ref{sec:intervention} further develops the case for this structural intervention.

\section{Multi-Persona Interfaces as a Critical Design Intervention}
\label{sec:intervention}

\begin{figure}[htbp]
  \centering
  \includegraphics[width=\linewidth]{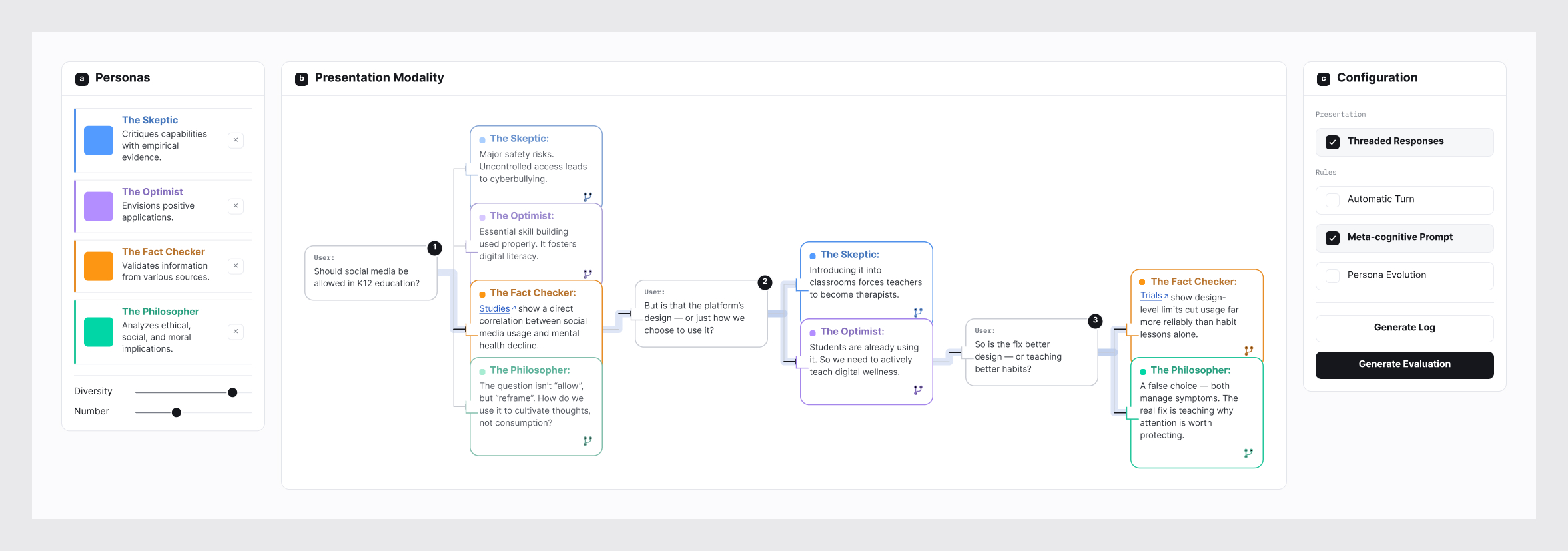}
  \caption{\textbf{Multi-Persona Conversational Interface Design Mockup.} 
    Instead of a single assistant voice, the user interacts with several different personas simultaneously. (a) A Skeptic, an Optimist, a Fact Checker, and a Philosopher, each with a stated role. The personas are adjustable for diversity and number. (b) The personas are designed to offer competing framings in parallel instead of a single combined answer, allowing the user to explore conversational branches across turns. Alternate reasoning paths are kept visible to the user. (c) Interface level controls allow the user to modify how plurality is presented and structured. The configuration shown corresponds to the facets of the design space developed in Section~\ref{subsec:design-space}. }
  
  \label{fig:mockup}
\end{figure}

This section proposes multi-persona interfaces as a structural response to the harm identified in the previous section (Figure~\ref{fig:mockup}).

\subsection{From Diagnosis to Intervention}
\label{subsec:diagnosis}

\citet{ghosh2026chatbots} argue that the chatbot is a dominant sociotechnical
configuration whose design choices (single authoritative responses, opaque
reasoning, cooperative Q\&A framing) inherently reduce user agency, and propose
four intervention strategies, among them redesigning chatbots to present multiple
perspectives rather than single answers. This paper extends that point into a
sustained argument: multiple perspectives presented at the interface could
simultaneously scaffold metacognitive reflection, expose users to ideological
diversity, and surface the pluralistic values monolithic alignment suppresses.

A key driver of cognitive offloading in standard chatbot interactions is the low
inferential distance between the model's output and an apparently finished answer
\citep{anderson2024homogenization}. When a model produces a polished,
complete-seeming response, why would a user interrogate it further? Multi-persona
interfaces can counter this dynamic by systematically refusing to provide a
single answer, instead presenting the user with perspectives that must be
actively evaluated and comprehended. The interface itself does the work of
preserving plurality where the model's underlying training and alignment have
flattened it.

This intervention works on all three facets at once: cognitively, plural
presentation restores the inferential demand that single-voice interfaces
eliminate; ideologically, it makes the contestability of contested topics
visible; epistemically, it creates space for marginalized perspectives and
knowledge systems to appear alongside dominant ones.

\subsection{Why Plurality Works: Lessons from Game Design}

Multiplicity interface design in LLMs draws from three insights gleaned from conventions in multiplayer online game environments discussed in Section~\ref{subsec:design-history}. Each of these design patterns helps explain why plurality design produces a value-laden understanding than a single, authoritative answer.

The first feature is the frame. \citet{huizinga1955homoludens} posits the “magic circle”, which is formalized into game design theory by \citet{tekinbas2003rules}. It further establishes a boundary between the game world and ordinary life. Inside, rules in ordinary life are temporarily suspended in order to carry out the acts of play, non-functional, voluntary, and  intrinsically rewarding acts of exploration. \citet{castronova2005synthetic} further introduces the concept of the porous membrane. Rather than an absolutely isolated virtual world, Castronova defines MMORPG as a permeable space that  "protect[s] the inner world with its set of rules, and yet at the same time allow[s] players to cross the barrier, bringing their real-life assumptions to the game." This porosity bestows MMORPG worlds with their epistemic value. In this setting, dissent is not only more affordable, but carries implications to activities taking place in the physical reality as well. Similarly, by introducing synthetic voices while allowing the user to speak as herself, multiplicity interface design follows the porous membrane principle as well. While a single-assistent chat interface frames the user as a requester waiting for a single, completed answer, plurality chat interface can frame the user as a participant in an exploratory deliberation. 

The second feature is role differentiation. In multiplayer video games, role assignment is largely structural, giving the player a vantage point that stems from their position within the group dynamic, rather than personal taste or preference. For example, in a raid, the tank engages with the NPC boss’s attacks and absorbs its damage, the healer manages group health, and the mage performs specialty actions such as crowd control and burst damage. All views are simultaneously legible yet partial, structurally mandating that all players exchange purview, synthesize information, and work in tandem to spread knowledge across the entire guild. The design and distribution of expert roles grants dissent a warrant, and discourages group-level conformity. Plurality LLM interface design carries over this feature by introducing various epistemic roles to ensure diverse perspectives and values. However, there is one major difference. In a group-based game, the guild as a whole synthesizes the views. In plurality interface design, the user alone performs the synthesis, because the rest of the voices are procedurally generated without human intervention. 

The third feature is the asymmetric self. The roles are fictional, but the cause and effects are consequential. Consider a guild raid, a scheduled, high-stakes multiplayer event where a group of players collaborates to defeat procedurally generated bosses. Because of their difficulty and complexity, raids call for meticulous organization, coordination, and resource commitment to succeed. The consequence of failure is real. However, because all guild members’ personas are fictional, temporarily assigned, and can be changed fluidly, the pressure from a real-world discourse group is partially alleviated. Therefore, MMORPG interface design is situated somewhere between two ends of the spectrum: real-world interpersonal exchanges, which are sometimes too rigid to inspire experimental exchanges, and pure roleplay environments, where the consequences are too loosely defined to support real epistemological efforts. The multi-persona interface also rests on an asymmetric design, but with a difference. The personas are fictional, but the users stay as themselves throughout the deliberation, which allows users to balance between exploration and accountability. 

Taken together, these three features provide the design logic for the pluralistic interface. The frame makes disagreement safe, role differentiation ensures exchange of perspectives, while the asymmetric self keeps its consequence relevant for the user. 

\subsection{Precursors in Conversational AI Design}
\label{subsec:precursors}

Existing work in conversational AI design has begun to demonstrate that interface-level changes can meaningfully shift user cognitive engagement. \citet{favero2025socratic} designed a chatbot to mimic Socratic questioning, bypassing direct answers and minimizing validation in favour of probing questions; simulated students using it significantly outperformed those using a typical chatbot. \citet{xi2026socratic} corroborate these results with their Socratic Intelligent Conversational Agent, designed for similar purposes and tested with 94 university students, finding that students using the Socratic agent obtained higher scores on their research projects and significantly higher reflective thinking scores. Interview anecdotes revealed that participants in the experimental condition felt that the Socratic chatbot mitigated cognitive dependence and degradation typically associated with chatbot tools.

\citet{singh2025metacognitive} tested a similar concept in a different domain,
investigating the impact of deliberately designed metacognitive prompts on
critical thinking during generative-AI-supported search tasks. Prompts directing
attention to one's own thought processes at strategic intervals increased
cognitive engagement relative to generative-AI search without them, evidenced by
broader topic exploration and more critical questions. Outside the classroom,
\citet{lee2025catalysts} find that a Socratic agent providing critical feedback
during collaborative design decisions increased satisfaction with teamwork,
process, and outcome.

These case studies foreground a design principle that our multi-persona interface emphasizes: they do not display a single, definitive answer to any given conversation topic. However, they are still single-voice interface designs. Productive cognitive scaffolding is introduced, in this case, through the reflective structure embedded in the conversation's progression, rather than the presentation of multiple perspectives.

ChatGraPhT \citep{kimm2025chatgrapht} moves closer to
multi-perspective presentation with a branching node-link interface: users branch
conversations into multiple paths, merge ideas across branches, and receive
guidance from two agentic assistants, making the conversation's structure
visible. Though evaluated through author-based inquiry rather than a formal user
study, it illustrates how making alternative reasoning paths explicit can produce
cognitive engagement that standard interfaces suppress. \citet{ye2026reflexis}
take a more structured approach with a Human-AI deliberation framework, in which
users are required to commit to a position before being shown AI-generated
counterarguments, with the goal of preserving the user's epistemic autonomy
across the interaction.

Several studies have begun to test the effects of exposure to multiple AI voices
simultaneously. Both studies below employ multi-agent architectures as the
vehicle for diverse perspectives; in the taxonomy of
Section~\ref{subsec:taxonomy} their user-facing function is multi-persona while
their substrate is multi-agent, and their findings support the multi-persona
design principle either way. \citet{song2025multiagents} investigated the social
influence effects of groups of AI agents on users' opinions about contested
topics such as self-driving cars and violent video games. They found that
exposure to multiple AI agents with distinct stances shifted users' opinions more
than a single agent, and with younger users and those with lower education more
susceptible. This finding demonstrates that presenting users with multiple
distinct perspectives --- here implemented through a multi-agent architecture ---
can diversify the perspectives users encounter. It also indicates a risk we
return to in Section~\ref{subsec:risks}: when the persona group is itself
homogeneous or unbalanced, multi-persona presentation can amplify rather than
mitigate conformity pressure.

Evidence from the creativity domain reinforces this point.
\citet{lu2024discussion} found that a multi-agent discussion framework, in which
LLM agents with distinct role-play personas engage in structured three-phase
exchanges, outperformed on originality across all four benchmarks and on
elaboration across three of four. Notably, a single agent cycling through
multiple personas underperformed the multi-agent version, suggesting that genuine
inter-agent exchange, not merely the presence of diverse labels, is what drives
divergence in outputs.

\subsection{Pluralistic Alignment as a Technical Foundation}
\label{subsec:alignment}

A parallel body of work, often gathered under the heading of pluralistic
alignment, has developed technical methods for producing AI systems whose outputs
reflect a broader range of human values, perspectives, and cultural contexts than
the standard alignment pipeline tends to produce. \citet{sorensen2024roadmap}
provide an influential roadmap, distinguishing three forms of pluralistic
alignment: presenting users with Overton-window-spanning sets of acceptable
responses, steerable systems that can adopt particular value profiles on demand,
and distributional pluralism in which model outputs reflect the empirical
distribution of human views.

A growing body of technical work instantiates these directions. Modular pluralism
proposes a framework in which a base LLM collaborates with a pool of smaller
community language models, each fine-tuned on politically or culturally distinct
corpora \citep{feng2024modular}. Depending on the task, the base model either
summarizes diverse community inputs, selects the most relevant community voice,
or aggregates probability distributions across communities. VISPA
\citep{zheng2026vispa}, EthosAgents \citep{zhong2025healthcare}, Cultural Palette
\citep{yuan2026palette}, and PluralLLM \citep{srewa2025plurallm} each take
different technical approaches to producing model outputs that reflect plural
values rather than a smoothed-over consensus. These systems demonstrate that
pluralistic outputs are technically achievable and that the alignment process can
be rebuilt to surface, rather than collapse, plurality at the model layer.

Multi-persona interfaces can be built on top of pluralistically aligned models, drawing on the diverse outputs the model produces and presenting them to the user as distinct, contestable voices. Conversely, pluralist alignment at the model level, without appropriate interface design, risks flattening diverse output back into a single voice when communicating it to end users. To avoid undermining the effort of model-level diversity, technical and interface design interventions must be considered in tandem. 

\subsection{The Design Space of Multi-Persona Interfaces}
\label{subsec:design-space}

A serious UX intervention has a design space: the parameters and constraints that
define what the system can be. Articulating it prepares for both implementation
and empirical study. We propose the following starting taxonomy
(Figure~\ref{fig:design-space}), drawing on the precursors in
Section~\ref{subsec:precursors}, the genealogies in
Section~\ref{subsec:design-history}, and our own design intuitions.

\textbf{Number of Personas:} Without external aid or rehearsal techniques, users have limited working memory. Consequently, the number of active voices presented via a digital user interface shapes the player's cognitive load during gameplay. Party-based role-playing games where players control a group of characters typically present three to five active in-game voices to maintain a high level of player engagement. While MMORPGs track thousands of active players at any given time, it also uses a clustered design principle to group information into a limited number of categories, or chunks. For example, in EVE Online, instead of tracking conversation lines with thousands of ally members, the system compresses the battle information down into three to five focal points: a single guild leader who oversees the battle of 500 guild members; a single enemy stat caller that aggregates 500 enemy targets into 1 consolidated target; a single visual bar representing group health stats; a single audio alert to flankers; a single personal health stat bar. 

\textbf{Persona Diversity:} When multiple personas share core values but differ only in minor cosmetic details, such as tone or phrasing, the system risks reinforcing the user's existing bias. Conversely, when personas hold wildly differing opinions and lack a shared underlying narrative context, the user cannot determine group dynamics or evaluate relative trade-offs. Users default to arbitrary behaviors to reduce cognitive overload. Multiplayer game design offers case studies such as Space Station 13 and Barotrauma, where players are assigned functional personas such as Captain, Engineer, Geneticist, and Security Officer. Not a single player has full access to the overall station design. They must learn to communicate what they know and work together as a collective.

\textbf{Presentation Modality:}  How personas are shown affects the pattern of engagement they produce. Side-by-side parallel display, sequential turn-taking, hierarchical arrangement (one primary voice with dissent available on demand) each carry different cognitive demands.

\textbf{User Control Over Persona Configuration:} Can the user select which
personas to engage, modify them, or introduce new ones? High-control
configurations approach the participatory worldbuilding of zzyw's
\emph{ThingThingThing} \citeyearpar{zzyw2019thingthingthing}; low-control
configurations resemble curated newsstands, the user choosing among sets the
designers prepared. A moddable interface lets users treat agent logic as an
avatar, altering the AI's persona for identity-driven tasks.

\textbf{Temporal Dynamics:} Persona configurations may be static or dynamic, with personas introduced or retired as the conversation progresses, or the set evolving with user engagement. Game design has a rich precedent for dynamic character introduction and arc development that AI interface design has not absorbed; personas might evolve to challenge the user's growing consensus, sustaining productive friction throughout.

\textbf{Persona Attribution:} How each voice is marked shapes willingness to
engage with it. Named characters, color-coding, ideological labels, demographic
descriptors, role descriptors (expert, peer, skeptic), and institutional voices
carry different connotations. Drawing on the ``Proteus Effect''
\citep{yee2014proteus}, whether a persona reads as peer, rival, or guild-mate
directly influences willingness to enter critical debate. The attribution scheme
is itself a site where the interface's value commitments become visible.

\textbf{Persona Weighting and Sequencing:} Not all contributions are equally salient at every moment. MMORPG raiding has an implicit sequencing logic: the tank engages first, the healer monitors continuously, the damage dealers execute within the space the others create. Deliberation may have analogous phases, with expansive frames more productive early and critical ones later. Whether the interface should encode this or leave it to the user is open: imposed sequencing risks rigidity, purely user-directed sequencing risks the avoidance patterns the design is meant to counteract.

\textbf{Persona Roster Definition:}  In MMORPGs, role structures are co-produced by designers, encounter constraints, and accumulated community practice; no single authority defines what roles must exist. The multi-persona interface has no equivalent mechanism. A user-defined roster gives full compositional agency but risks selection bias, users inviting only friction they find comfortable. A system-suggested roster raises questions about who classifies problem types. A community-generated roster, analogous to game communities developing role meta-strategies, requires a user base and feedback infrastructure that does not yet exist.

Cutting across these dimensions is the balance between deliberative seriousness
and exploratory playfulness. Formal environments such as courtrooms and
parliamentary hearings produce adversarial engagement but discourage exploration,
because positions carry personal and institutional stakes. The game design
traditions in Section~\ref{subsec:design-history} suggest an alternative: bounded
spaces where stakes feel meaningful but exploring divergent positions costs
little. Multi-persona interfaces occupy this tension. Too serious, and users
default to defensive argumentation; too playful, and they disengage from the
substantive stakes. Persona attribution, presentation modality, and user control
collectively determine where a configuration falls on this spectrum, and getting
the balance right is an empirical question for Section~\ref{sec:agenda}.

These dimensions are illustrative rather than exhaustive. The actual design space will be richer, and the most productive work will come from collaboration between AI researchers, HCI designers, and the traditions in games, participatory simulation, and interactive art that have already worked through plural-interface problems.

\begin{figure}[htbp]
  \centering
\definecolor{ffourlo}{HTML}{DFE2EC}
\definecolor{ffourhi}{HTML}{2446BE}
\definecolor{ffourtext}{HTML}{1A1A1A}
\definecolor{ffourlabel}{HTML}{444444}
\definecolor{ffourgreen}{HTML}{4A9A35}
\providecommand{\ffourrow}[4]{%
  \begin{scope}[yshift={-(#1-1)*\ffourpitch}]
    \node[anchor=base west,inner sep=0pt,text=ffourtext] at (0,0)
      {\palheadface\bfseries\fontsize{7.2}{8.6}\selectfont #1\,\textperiodcentered\,#2};
    \shade[left color=ffourlo,right color=ffourhi,rounded corners=1.8pt]
      (\ffourxa,-10) rectangle (\ffourxb,-13.6);
    \node[anchor=base west,inner sep=0pt,text=ffourlabel] at (\ffourxa,-22)
      {\palheadface\fontsize{6}{7}\selectfont #3};
    \node[anchor=base east,inner sep=0pt,text=ffourlabel] at (\ffourxb,-22)
      {\palheadface\fontsize{6}{7}\selectfont #4};
  \end{scope}}
\begin{tikzpicture}[x=\linewidth,y=1pt]
  \def\ffourxa{0.155}\def\ffourxb{0.875}\def\ffourpitch{38}
  \node[anchor=base west,inner sep=0pt,text=ffourtext] at (0,48)
    {\palheadface\bfseries\fontsize{10}{12}\selectfont
     The design space of multi-persona interfaces};
  \node[anchor=base west,inner sep=0pt,text=palmuted,text width=\linewidth,
        align=left] at (0,36)
    {\raggedright\palheadface\fontsize{6.2}{8}\selectfont
     \spaceskip=\fontdimen2\font\xspaceskip=\spaceskip
     Eight dimensions along which any multi-persona configuration can be
     located. Each axis runs from a simpler pole to a richer one.\par};
  \node[anchor=base west,inner sep=0pt] at (\ffourxa,14)
    {\pallabel[palmuted!75]{Simpler pole}};
  \node[anchor=base east,inner sep=0pt] at (\ffourxb,14)
    {\pallabel[palmuted!75]{Richer pole}};
  \ffourrow{1}{Number of personas}{two (for / against)}
    {party-sized $\rightarrow$ many co-present voices}
  \ffourrow{2}{Similarity \& diversity}{clustered / homogeneous}
    {wide spread \textperiodcentered{} \textquotedblleft honest social friction\textquotedblright}
  \ffourrow{3}{Presentation modality}{sequential turn-taking}
    {side-by-side \textperiodcentered{} hierarchical \textperiodcentered{} hybrid}
  \ffourrow{4}{User control}{curated newsstand}
    {fully moddable / user-authored}
  \ffourrow{5}{Temporal dynamics}{static roster}
    {personas appear, retire, evolve mid-session}
  \ffourrow{6}{Persona attribution}{abstract labels / colors}
    {named roles \textperiodcentered{} demographic \textperiodcentered{} institutional}
  \ffourrow{7}{Weighting \& sequencing}{flat / user-directed}
    {phase-aware system sequencing}
  \ffourrow{8}{Roster definition}{user-defined}
    {system-suggested $\rightarrow$ community-generated}
  \draw[ffourgreen,line width=1pt,line cap=round,dash pattern=on 0pt off 2.6pt]
    (0.925,4) -- (0.925,{-7*\ffourpitch-20});
  \node[rotate=-90,anchor=center,inner sep=0pt,text=ffourgreen]
    at (0.958,{-3.5*\ffourpitch-8})
    {\palheadface\bfseries\fontsize{6.4}{7.6}\selectfont
     cross-cutting axis: deliberative seriousness $\leftrightarrow$ exploratory playfulness};
\end{tikzpicture}
  \caption{\textbf{Any multi-persona configuration can be located along eight dimensions,
  each running from a simpler pole (left) to a richer one (right):} the number of
  personas presented, the diversity of perspectives across them, the modality of
  presentation, the degree of user control over configuration, the temporal
  dynamics of the persona set, the scheme used to attribute each voice, the
  weighting and sequencing of contributions, and the source of the persona
  roster. A ninth consideration shown along the right edge, the balance between
  deliberative seriousness and exploratory playfulness, is not set independently
  but emerges from choices across all eight dimensions.}
  \label{fig:design-space}
\end{figure}

\subsection{Risks, Design Considerations, and the Need for Careful Evaluation}
\label{subsec:risks}

Multi-persona interfaces are not risk-free, and the case for them depends on
attending to several risks that the single-persona configuration does not present
in the same form.

\textbf{Manufactured Conformity:} Song et al.'s
\citeyearpar{song2025multiagents} social influence study discussed in
Section~\ref{subsec:precursors} found that exposure to homogeneous groups of AI
agents shifted user opinion in the direction of the group. A multi-persona
interface in which the persona configuration is uniform or biased can amplify,
rather than mitigate, the ideological narrowing of single-persona interfaces. The
intervention's value depends on genuine diversity in the persona configuration.

\textbf{Manipulation at Scale:} Persuasion research \citep{bai2025persuade,
schoenegger2026persuasive} shows that chatbots are highly effective persuaders,
often more so than humans on similar tasks. A multi-persona interface that
presents apparently plural voices but is engineered to nudge users toward a
particular conclusion would be a more sophisticated form of manipulation than the
current single-persona configuration. The interface's plurality must be
substantive rather than theatrical, and design choices around persona attribution
and configuration are sites where this distinction becomes visible or hidden.

\textbf{User Cognitive Load:} The single-persona interface succeeds in part
because it minimizes user cognitive demand. Multi-persona interfaces deliberately
reintroduce that demand. For some users and tasks this is desired; for others it
may produce fatigue or disengagement, so different configurations will suit
different contexts.

\textbf{Sociocultural Context Dependence:} What counts as a meaningful diversity
of personas is itself culturally specific. A persona configuration that reads as
productively plural in one context may read as missing key voices in another.
Multi-persona interfaces, like other AI systems, will need ongoing localization
and contestation.

\textbf{Persona Fatigue:} One underexplored risk is the gradual erosion of a
persona's friction-generating effectiveness as users learn to anticipate and
discount its contributions. In MMORPGs a role's credibility is underwritten by
real consequences of failure; in the AI interface, consequences are real only
insofar as the user treats the deliberation as consequential. Whether and how
quickly users habituate to persona-based challenge, and what design features
sustain genuine opposition over repeated use, is unknown.

These risks call for careful attention to design ethics and to empirical study of
how users actually engage with these interfaces. The pluralistic alignment work
surveyed in Section~\ref{subsec:alignment} and the prototypes in
Section~\ref{subsec:precursors} have largely been evaluated through automated
metrics or expert annotation rather than through actual users over time. We know
pluralistic outputs can be produced; we know very little about what users do with
them in sustained use, what practices and social formations emerge, and how the
taxonomy's design decisions affect outcomes. Closing that gap is the focus of
Section~\ref{sec:agenda}.

\section{A Research Agenda for Plural-Interface AI}
\label{sec:agenda}

\subsection{Toward Careful Evaluation}

The highest-priority research direction we propose is to close the evaluation gap laid out in Section~\ref{subsec:risks}. This requires empirical work in several dimensions, including user studies across all facets of the design space described in section 5.5 that investigate how changing the number, diversity, and presentation of personas affects users’ cognitive engagement, ideological openness and value exposure. Single-shot studies alone are unlikely to reveal the full harms of single-persona interfaces, and how these harms accumulate over repeated use, so longitudinal study is also required. Equally important is evaluation on a heterogeneous population, as \citet{song2025multiagents} found that younger users and those with lower levels of formal education were more susceptible to chatbot influence. Lastly, a direct comparison against single-persona baselines is essential. This should go beyond the interface producing diverse output, and assess whether users engage with outputs in ways that meaningfully mitigate the previously discussed harms. 

Additional questions include the optimal number of personas, whether this number generalises across user groups, whether personas should evolve or change as the user’s deliberation developers, how persona definition affects epistemic coverage, and whether the multi-persona interface is substantially more difficult for users to engage with. In summary, the longitudinal impacts on users of interacting with multi-persona interfaces connect to broader questions about sustained human-AI interaction.

\subsection{The Sociology of Plural Interfaces}

As discussed in Section~\ref{subsec:design-history}, the traditions of game design and participatory simulation encourage the emergence of communities that prioritize interpretation and collective meaning-making over the seeking of definitive answers. This is a dimension that is typically overlooked in human-AI research on chatbots, but provides a unique lens through which to view the design of chatbots. 

It may be possible for multi-persona interfaces to similarly encourage the formation of such communities and ecosystems. Much like the behaviour of fan communities for character-centric games and tabletop roleplaying systems, as well as the dynamics of participatory simulations, chabot users may begin to share preferred persona configurations, develop categorisations of effective and ineffective persona types, contest the choices made by system designers, create their own personas and modify existing ones, or form sub-communities for specific, contested topics \citep{steinkuehler2006thirdplaces, boellstorff2008secondlife,
pearce2009communities}.

This illuminates several promising research directions. Firstly, studying the usage habits of the early adopters of multi-persona interfaces may reveal insights about the formation and evolution of user practices. Secondly, analysing the differences and similarities between multi-persona chatbots and game dynamics may yield insights about how to facilitate community building amongst users. Finally, investigating how users contest the values embedded into personas can reveal how chatbot interface design should evolve to support the needs of the users (for example, through modifiability and extensibility). 

This is where the AI social science perspective is most distinctive: where HCI
studies individual users in the lab and computer science studies model behavior
in the abstract, AI social science can study how user communities co-evolve with
the interfaces they engage.

\subsection{Pluralism in Practice: Cross-Cultural and Decolonial Considerations}

It is also important to consider cultural context: a multi-persona interface that operates well in the context of one community may operate poorly in another. The taxonomy discussed in Section~\ref{subsec:design-space} embeds the following assumptions: that persona voices can be distinct and individually attributable, that contestation is productive, and that plurality is a productive end-goal for chatbot interface design. Future work building on this paper should engage with how interfaces can be appropriately situated within, and rebuilt for, different cultural and institutional contexts. 

Research in this area connects to broader work on epistemic justice and the decolonization of AI \citep{ricaurte2019data, mhlambi2020ubuntu, birhane2021injustice}., which requires intentional co-design alongside communities that have been historically underrepresented in AI development. The taxonomy outlined in this paper should be tested with empirical studies of how multi-persona interfaces may support, or fail to support, non-Western knowledge and value systems, including indigenous, communal and tradition-based epistemologies. Additionally, multi-persona interface may reproduce hegemonic assumptions about which perspectives count as ‘diverse’, and which framings are considered default versus marginal, motivating further investigation.

This argument can be extended to the governance documents that increasingly define model behaviour. Currently, a single constitution or model spec encodes one value profile into a technology that’s used by a diverse global population \citep{openai2025modelspec, anthropic2026constitution}. Whilst a multi-persona interface could easily integrate several culturally-situated value documents, questions emerge regarding who authors them, and by what qualifications. Ultimately, pluralism, as we have used the term throughout this paper, is an intentional commitment that rests on the design and research choices made during its implementation.

\section{Conclusion}
\label{sec:conclusion}

We have argued in this paper that the current design paradigm of LLM-chatbots prioritizes a single, authoritative voice presenting seemingly complete answers, embedding an implicit sociotechnical commitment. Drawing on literature from cognitive science, political studies, AI alignment, game design and human computer interaction, we have demonstrated that this design commitment creates three related harms: 1) the offloading of productive cognitive load and critical thinking, 2) the narrowing of ideologies and beliefs, and 3) epistemic compression of pluralistic cultures and values. We then presented multi-persona interfaces as a viable intervention for addressing all three harms at once, proposing that these interfaces provide a relatively cheap, non-technical way to refuse the single-voice commitment. Finally, the paper outlined the design space, and laid out an agenda for future research that treats both the user, and the broader community of users, as the key units of evaluation. 


This paper finds a natural home in the AI and social science framing of this collection. We treat interface design as a sociotechnical configuration with both technical and social effects on the users’ cognition, beliefs, and knowledge of users. Rather than uphold the optimism-vs-pessimism binary, we adopt a position of thoughtful criticism by arguing that the dominant design paradigm produces specific harms, that alternatives exist, and that these alternatives may carry their own risks and should therefore be evaluated empirically.

Ultimately, the future of multi-persona interfaces relies on the willingness of researchers, designers, and user communities to do the work of building, deploying, evaluating and contesting them. The idea of pluralism-as-a-design-principle is empty without meaningful pluralism in the work that attempts to build it. Rather than settle, definitely, what pluralistic conversational AI should look like, this agenda aims to open the question to inquiry that is sustained, cross-disciplinary, and user-centered.

\section*{Data Availability}
Data sharing is not applicable to this article as no datasets were generated or
analyzed during the current study.

\section*{Ethics Declaration}
Ethical approval was not required as no human participants were involved in this
research.

\section*{AI Use Disclosure}
During the preparation of this work, the authors used Claude Opus in order to source literature, structure arguments, and perform edits for clarity. After using this tool, the authors reviewed and edited the content as needed. The authors take full responsibility for the content of the publication.

\bibliography{references}

\end{document}